\documentclass[12pt]{amsart}
\usepackage{amsmath,amssymb,color}
\usepackage{graphicx}
\usepackage{pgfnodes}
 \DeclareGraphicsExtensions{.jpg,.gif,.png,.pdf}
\newcommand{\<}{\langle}
\renewcommand{\>}{\rangle}
\newcommand{\uk}{|\kern-3.3pt\uparrow\>}
\newcommand{\dk}{|\kern-3.3pt\downarrow\>}
\newcommand{\ub}{\<\uparrow\kern-3.3pt|}
\newcommand{\db}{\<\downarrow\kern-3.3pt|}

\newcommand{\re}{\text{Re~}}
\newcommand{\im}{\text{Im~}}

\newcommand{\ds}{\displaystyle}

\title[The quantum walk of F. Riesz]{The quantum walk of F. Riesz}
\author[GV]{F.~A.~Gr\"unbaum, L.~Vel\'{a}zquez}
\thanks{The research of the first author was supported in part by
the Applied Math Sciences subprogram of the Office of Energy Research,USDOE, under
contract DE-AC03-76SF00098.}
\thanks{The research of the second author was partly supported
by the Spanish grants from
the Ministry of Education and Science, project code MTM2005-08648-C02-01, and
the Ministry of Science and Innovation, project code MTM2008-06689-C02-01,
and by Project E-64 of Diputaci\'on General de Arag\'on (Spain).}
\date{}
\address[F.~A.~Gr\"unbaum]{Department of Mathematics \\ University of
California \\ Berkeley \\ CA \\ 94720}
\address[L.~Vel\'{a}zquez]{Departamento de
Matem\'{a}tica Aplicada \\ Universidad de Zaragoza \\ Zaragoza \\
Spain}
\subjclass[2000]{81P68, 47B36, 42C05}
\keywords{Riesz measure, Laurent orthogonal polynomials,
CMV matrices, quantum random walks}

\begin{document}

\begin{abstract}
We exhibit a way to associate a quantum walk (QW) on the non-negative integers
to any probability measure on the unit circle. This forces us to
consider one step transitions that are not traditionally allowed. We illustrate
this in the case of a very interesting measure, originally proposed by F. Riesz for a different purpose.

For a review of Riesz's construction and its many uses, see \cite{Zy,Si04-1,Ka}.
For
reviews of quantum walks, see \cite{A,K,Ko}.

\bigskip

\end{abstract}

\maketitle

\section{Introduction and contents of the paper} \label{INT}

The purpose of this note is to consider the probability measure
constructed by F. Riesz, \cite{Ri}, back in 1918, and to study a
quantum walk naturally associated to it.

The measure on the unit circle that F. Riesz built is formally given
by the expression
\begin{equation} \label{RIESZ}
\begin{aligned}
d\mu(z) &= \prod_{k=1}^{\infty} (1+ \cos(4^{k} \theta))
\frac{d\theta}{2\pi}
= \prod_{k=1}^{\infty} (1+(z^{4^{k}}+z^{-4^{k}})/2) \frac{dz}{2\pi iz} \\
&= \sum_{j=-\infty}^{\infty}\overline\mu_j z^j \frac{dz}{2\pi iz}\,.
\end{aligned}
\end{equation}
Here $z=e^{i \theta}$. If one truncates this infinite product the
corresponding measure has a nice density. These approximations
converge weakly to the Riesz measure.

The recent paper \cite{CGMV} gives a natural path to associate to a
quantum walk on the non-negative integers a probability measure on
the unit circle. This construction is also pushed to quantum walks
on the integers. The traditional class of coined quantum walks
considered in the literature allows for certain one step transitions
and this leads to a restricted class of probability measures.

In this paper we take the attitude that for an arbitrary probability
measure a slightly more general recipe for these transitions gives
rise to a quantum walk. The measure considered by Riesz falls
outside of the more restricted class considered so far, and is used
here as an interesting example.

There is an obvious danger that having gone beyond the traditional
class of quantum walks some of the appealing properties of these
walks may no longer hold. From this perspective we consider the
example discussed here as a laboratory situation where we will test
some of these features.

There is an extra reason for looking at this special example:
Riesz's measure is one of the nicest examples of purely singular
continuous measures. This means that the unitary operator governing
the evolution of the corresponding quantum walk has a pure singular
continuous spectrum. Hence, Riesz's quantum walk becomes an ideal
candidate to analyze the dynamical consequences of such a kind of
elusive spectrum.

We first show how to introduce a quantum walk given a probability
measure on the unit circle and then we analyze in more detail the
case of Riesz's measure, and give some exploratory results
pertaining to the large time behaviour of the ``site distribution''
for this non-standard walk.

We are grateful to Prof. Reinhard Werner for pointing out that
F.~Riesz actually started the infinite product (\ref{RIESZ}) with
$k=0.$ There are two well known references \cite{Si04-1,GMcG} that
use the convention used here. Each choice has its own advantages as
will be seen in section 5.

For a review of Riesz's construction and its many uses, see
\cite{Zy,Si04-1,Ka}. For reviews of quantum walks, see
\cite{A,K,Ko}.

\bigskip

This paper will appear in the Proceedings of FoCAM 2011 held in Budapest, Hungary, to be published in the London Mathematical Society lecture Note Series.

\section{Szeg\H o polynomials and CMV matrices} \label{CMV}

Let $\mu$ be a probability measure on the unit circle
$\mathbb{T}=\{z\in\mathbb{C}:|z|=1\}$, and $L^2_\mu(\mathbb{T})$ the
Hilbert space of $\mu$-square-integrable functions with inner
product
\[
(f,g) = \int_\mathbb{T} \overline{f(z)}\,g(z)\,d\mu(z).
\]
For simplicity we assume that the support of $\mu$ contains an
infinite number of points.

A very natural operator to consider in our Hilbert space is given by
\begin{equation} \label{UMO}
U_\mu \colon \mathop{L^2_\mu(\mathbb{T}) \to L^2_\mu(\mathbb{T})}
\limits_{\ds \, f(z) \longrightarrow zf(z).}
\end{equation}
Since the Laurent polynomials are dense in $L^2_\mu(\mathbb{T})$, a
natural basis to obtain a matrix representation of $U_\mu$ is given
by the Laurent polynomials $\{\chi_j\}_{j=0}^\infty$ obtained from
the Gram--Schmidt orthonormalization of
$\{1,z,z^{-1},z^2,z^{-2},\dots\}$ in $L^2_\mu(\mathbb{T})$.

The matrix $\mathcal{C}=(\chi_j,z\chi_k)_{j,k=0}^\infty$ of $U_\mu$
with respect to $\{\chi_j\}$ has the form
\begin{equation} \label{C}
\mathcal{C} =
\begin{pmatrix}
\overline{\alpha}_0 & \kern-3pt \rho_0\overline{\alpha}_1 &
\kern-3pt \rho_0\rho_1 & \kern-3pt 0 & \kern-3pt 0 & \kern-3pt 0 &
\kern-3pt 0 & \kern-3pt \dots
\\
\rho_0 & \kern-3pt -\alpha_0\overline{\alpha}_1 & \kern-3pt
-\alpha_0\rho_1 & \kern-3pt 0 & \kern-3pt 0 & \kern-3pt 0 &
\kern-3pt 0 & \kern-3pt \dots
\\
0 & \kern-3pt \rho_1\overline{\alpha}_2 & \kern-3pt
-\alpha_1\overline{\alpha}_2 & \kern-3pt \rho_2\overline{\alpha}_3 &
\kern-3pt \rho_2\rho_3 & \kern-3pt 0 & \kern-3pt 0 & \kern-3pt \dots
\\
0 & \kern-3pt \rho_1\rho_2 &  \kern-3pt -\alpha_1\rho_2 & \kern-3pt
-\alpha_2\overline{\alpha}_3 & \kern-3pt -\alpha_2\rho_3 & \kern-3pt
0 & \kern-3pt 0 & \kern-3pt \dots
\\
0 & \kern-3pt 0 & \kern-3pt 0 & \kern-3pt \rho_3\overline{\alpha}_4
& \kern-3pt -\alpha_3\overline{\alpha}_4 & \kern-3pt
\rho_4\overline{\alpha}_5 & \kern-3pt \rho_4\rho_5 & \kern-3pt \dots
\\
0 & \kern-3pt 0 & \kern-3pt 0 & \kern-3pt \rho_3\rho_4 & \kern-3pt
-\alpha_3\rho_4 & \kern-3pt -\alpha_4\overline{\alpha}_5 & \kern-3pt
-\alpha_4\rho_5 & \kern-3pt \dots
\\
\dots & \kern-3pt \dots & \kern-3pt \dots & \kern-3pt \dots &
\kern-3pt \dots & \kern-3pt \dots & \kern-3pt \dots & \kern-3pt
\dots
\end{pmatrix},
\end{equation}
where $\rho_j=\sqrt{1-|\alpha_j|^2}$ and $\{\alpha_j\}_{j=0}^\infty$
is a sequence of complex numbers such that $|\alpha_j|<1$. The
coefficients $\alpha_j$ are known as the Verblunsky (or Schur, or
Szeg\H o, or reflection) parameters of the measure $\mu$, and
establish a bijection between the probability measures supported on
an infinite set of the unit circle and sequences of points in the
open unit disk. The unitary matrices of the form above are called
CMV matrices, see \cite{Si04-1,SiFIVE,Wa93}.

The problem of finding the sequence $\{\alpha_j\}$ for a given
measure $\mu$ or, more generally, that of relating properties of the
measure and the sequence is a central problem, see \cite{Si04-1},
where a few explicit examples are recorded. Even in cases when the
measure is a very natural one, this can be a hard problem. Back in
the 1980's one of us formulated a conjecture based on work on the
limited angle problem in X-ray tomography. The same conjecture was
also made in a slightly different context in work of Delsarte,
Janssen and deVries. The conjecture amounts to showing that the
Verblunsky parameters of a certain measure are all positive. This
was finally established in a real tour-de-force in \cite{Mag}.

One of the results of this paper consists of finding these
parameters in the case of F.~Riesz's measure. In the process of
finding these parameters we will need to invoke some other
sequences. Some of these will be subsequences of $\{\alpha_j\}$, and
some other ones will only have an auxiliary role. We will propose an
ansatz for the Verblunsky parameters of the Riesz measure that have
been checked so far for the first 6000 non-null Verblunsky
parameters. This is enough for computational purposes concerning the
related quantum walk. A proof of our ansatz deserves additional
efforts.

The decomposition of a measure $d\mu$ above into an absolutely
continuous and a singular part can be further refined by splitting
the singular part into point masses and a singular continuous part.
The example of Riesz that we will consider later will consist only
of this third type of measure, and it is (most likely) the first
known example of a measure of this kind, built in terms of a formal
Fourier series. For the case of the unit interval there is a
construction of such a singular continuous measure in the classical
book by F.~Riesz and B.~Sz-Nagy which is most likely due to
Lebesgue. Notice that the method of ``Riesz products" introduced in
\cite{Ri} can be used to produce measures such as the one that lives
in the Cantor middle-third set. However in this case, as well as in
the one due to Lebesgue, one loses the tight connection with Fourier
analysis that makes the example of Riesz easier to handle.

A very important role will be played by the Carath\'{e}odory
function $F$ of the orthogonality measure $\mu$, defined by
\begin{equation} \label{CF}
F(z) = \int_\mathbb{T} \frac{t+z}{t-z}\,d\mu(t), \qquad |z|<1.
\end{equation}
$F$ is analytic on the open unit disc with McLaurin series
\begin{equation} \label{MOM}
F(z) = 1 + 2\sum_{j=1}^\infty\overline\mu_jz^j, \qquad \mu_j =
\int_\mathbb{T} z^jd\mu(z),
\end{equation}
whose coefficients provide the moments $\mu_j$ of the measure $\mu$.

Another useful tool in the theory of orthogonal polynomials on the
unit circle is the so called Schur function related to $\mu$ by
means of $F$ through the expression
\[
f(z) = z^{-1} (F(z)-1)(F(z)+1)^{-1}, \quad |z|<1.
\]
Since the Schur function, and its Taylor coefficients will play such
an important role, see \cite{GVWW}, we will settle the issue of
names by sticking to the name Verblunsky parameters for those that
could be also called by the names of Schur or Szeg\H o.

These functions obtained here by starting from a probability measure
on $\mathbb{T}$ can be characterized as the analytic functions on
the unit disk $\mathbb{D}=\{z\in \mathbb{C} :|z|<1\}$ such that
$F(0)=1$, $\re F(z)>0$ and $|f(z)|<1$ for $z\in\mathbb{D}$,
respectively.

Starting from $f_0=f$, the Verblunsky parameters $\alpha_k=f_k(0)$
can be obtained through the Schur algorithm that produces a sequence
of functions $\{f_k\}$ by means of
\begin{equation} \label{E:SchurAlg-scalar}
f_{k+1}(z) = \frac{1}{z}
\frac{f_k(z)-\alpha_k}{1-\overline{\alpha}_kf_k(z)}\,.
\end{equation}
By using the reverse recursion
\begin{equation}
f_k(z) =
\frac{zf_{k+1}(z)+\alpha_k}{1+\overline{\alpha}_kzf_{k+1}(z)} =
\alpha_k + \frac {\rho^2_k}{\overline{\alpha}_k + \frac
{1}{zf_{k+1}(z)}}
\end{equation}
one can obtain a continued fraction expansion for $f(z)$. This is
called a ``continued fraction-like'' algorithm by Schur, \cite{Sch},
and made into an actual one by H.~Wall in \cite{Wa}. See also
\cite{Si04-1}. We will illustrate the power of this way of computing
these parameters by using it in our example to compute (with
computer assistance, in exact arithmetic) enough of them so that we
can formulate an ansatz as to the form of these parameters.

\section{Traditional quantum walks} \label{RW}
We consider one-dimensional quantum walks with basic states
$|i\>\otimes\uk$ and $|i\>\otimes\dk$, where $i$ runs over the
non-negative integers, and with a one step transition mechanism
given by a unitary matrix $U$. This is usually done by considering a
coin at each site $i$, as we will see below.

One considers the following dynamics: a spin up can move to the
right and remain up or move to the left and change orientation. A
spin down can either go to the right and change orientation or go to
the left and remain down.

In other words, only the nearest neighbour transitions such that the
final spin (up/down) agrees with the direction of motion
(right/left) are allowed. This dynamics bears a resemblance to the
effect of a magnetic interaction on quantum system with spin: the
spin decides the direction of motion. This rule applies to values of
the site variable $i \ge 1$ and needs to be properly modified at
$i=0$ to get a unitary evolution.

Schematically, the allowed one step transitions are
\[
\begin{aligned}
|i\>\otimes\uk \longrightarrow \begin{cases} |i+1\>\otimes\uk &
\text{with amplitude} \hskip 5pt c_{11}^i
\\
|i-1\>\otimes\dk & \text{with amplitude} \hskip 5pt c_{21}^i
\end{cases}
\\
|i\>\otimes\dk \longrightarrow \begin{cases} |i+1\>\otimes\uk &
\text{with amplitude} \hskip 5pt c_{12}^i
\\
|i-1\>\otimes\dk & \text{with amplitude} \hskip 5pt c_{22}^i
\end{cases}
\end{aligned}
\]
where, in the case $i=0$, the unitarity requirement forces the
identification $|\kern-3pt-\kern-2pt1\>\otimes\dk \equiv
|0\>\otimes\uk$. For each $i=0,1,2,\dots$,
\begin{equation} \label{COIN}
C_i = \begin{pmatrix} c_{11}^i & c_{12}^i \\ c_{21}^i & c_{22}^i
\end{pmatrix}
\end{equation}
is an arbitrary unitary matrix which we will call the $i^{th}$ coin.

If we choose to order the basic states of our system as follows
\begin{equation} \label{ORD}
|0\>\otimes\uk, \; |0\>\otimes\dk, \; |1\>\otimes\uk, \;
|1\>\otimes\dk, \; |2\>\otimes\uk, \; |2\>\otimes\dk, \; \dots
\end{equation}
then the transition matrix is given below
\[
U = \begin{pmatrix}
c^0_{21} & 0 & c^0_{11} \\
c^0_{22} & 0 & c^0_{12} & 0 \\
0 & c^1_{21} & 0 & 0 & c^1_{11} \\
& c^1_{22} & 0 & 0 & c^1_{12} & 0 \\
& & 0 & c^2_{21} & 0 & 0 & c^2_{11} \\
& & & c^2_{22} & 0 & 0 & c^2_{12} & 0 \\
& & & & \ddots & \ddots & \ddots & \ddots & \ddots
\end{pmatrix}
\]
and we take this as the transition matrix for a traditional quantum
walk on the non-negative integers with arbitrary (unitary) coins
$C_i$ as in (\ref{COIN}) for $i=0,1,2,\dots$\,.

The reader will notice that the structure of this matrix is not too
different from a CMV matrix for which the odd Verblunsky parameters
vanish. This feature will guarantee that in the CMV matrix the
central $2 \times 2$ blocks would vanish identically. The CMV matrix
should have real and positive entries in some of the $2 \times 1$
matrices that are adjacent to the central $2 \times 2$ blocks, and
this is not generally true for the unitary matrix given above. In
\cite{CGMV} one proves that this can be taken care of by an
appropriate conjugation with a diagonal matrix. This associates a
measure $\mu$ on the unit circle with the above matrix $U$. Indeed,
$U$ becomes the matrix representation of the operator $U_\mu$,
defined in (\ref{UMO}), with respect to an orthonormal basis of
Laurent polynomials $X_j$ differing only by constant phase factors
$e^{i\theta_j}$ from the standard ones $\{\chi_j\}$ giving the CMV
matrix.

In \cite{CGMV} one considers the case of a constant coin $C_i$ for
which the measure $\mu$ and the function $F(z)$ are explicitly
found. In this case, after the conjugation alluded to above, the
Verblunsky parameters are given by
\begin{equation}
a,0,a,0,a,0,a,0\dots
\end{equation}
for a value of $a$ that depends on the coin, and the function $F(z)$
is, up to a rotation of the variable $z$, given by the function
\[
F(z) = -\frac{z-z^{-1}-2i\im a}{\sqrt{(z-z^{-1})^2+4|a|^2}-2\re
a}\,.
\]
The corresponding Schur function is the even function of $z$
\[
f(z) = \frac{z^2-1+\sqrt{(z^2-1)^2+4|a|^2z^2}}{2\overline{a}z^2}\,.
\]

It is easy to see that, in general, the condition $f(-z)=f(z)$ is
equivalent to requiring that the odd Verblunsky parameters of $\mu$
should vanish. Traditional coined quantum walks are therefore those
whose Schur function is an even function of $z$. In terms of the
Carath\'{e}odory function the restriction to a traditional quantum
walk amounts to $F(-z)F(z)=1$.

\section{Quantum walks resulting from an arbitrary probability measure}
One of the main points of \cite{CGMV} was to show that the use of
the measure $\mu$ allows one to associate with each state of our
quantum walk a complex valued function in $L^2_\mu(\mathbb{T})$ in
such a way that the transition amplitude between any two sates in
time $n$ is given by an integral with respect to $\mu$ involving the
corresponding functions and the quantity $z^n$. More explicitly we
have
\begin{equation} \label{E:KMcG-gen}
\< \tilde\Psi | {U}^n | \Psi \> = \int_\mathbb{T} z^n {\psi}(z)
\overline{\tilde{\psi}(z)}\ d{\mu}(z) ,
\end{equation}
where ${\psi}(z)=\sum_j\psi_jX_j(z)$ is the
$L^2_{{\mu}}(\mathbb{T})$ function associated with state
$|\Psi\>=\sum_j\psi_j|j\>$. Here $|j\>$ is the $j$-th vector of the
ordered basis consisting of basic vectors as given in (\ref{ORD}),
i.e. $|j\>$ stands for a site and a spin orientation, while $X_j(z)$
are the orthonormal Laurent polynomials related to the transition
matrix of the quantum walk. Similarly ${\tilde\psi}(z)$ is the
function associated to the state $|\Psi\>$.

This construction will now be extended to the case of any transition
mechanism that is cooked out of a CMV matrix as above. More
explicitly, we allow for the following dynamics
\[
\begin{aligned}
|i\>\otimes\uk \longrightarrow \begin{cases} |i+1\>\otimes\uk &
\text{with amplitude} \hskip 5pt \rho_{i+2} \rho_{i+3}
\\
|i-1\>\otimes\dk & \text{with amplitude} \hskip 5pt
\rho_{i+1}\overline{\alpha}_{i+2}
\\
|i\>\otimes\uk &\text{with amplitude} \hskip 5pt
-\alpha_{i+1}\overline{\alpha}_{i+2}
\\
|i\>\otimes\dk &\text{with amplitude} \hskip 5pt
\rho_{i+2}\overline{\alpha}_{i+3}
\end{cases}
\\
|i\>\otimes\dk \longrightarrow \begin{cases} |i+1\>\otimes\uk &
\text{with amplitude} \hskip 5pt -\alpha_{i+2} \rho_{i+3}
\\
|i-1\>\otimes\dk & \text{with amplitude} \hskip 5pt
\rho_{i+1}\rho_{i+2}
\\
|i\>\otimes\uk &\text{with amplitude} \hskip 5pt
-\alpha_{i+1}\rho_{i+2}
\\
|i\>\otimes\dk &\text{with amplitude} \hskip 5pt
-\alpha_{i+2}\overline{\alpha}_{i+3}.
\end{cases}
\end{aligned}
\]
The expressions for the amplitudes above are valid for any even $i$
with the convention $|\kern-2pt-\kern-2pt1\>\otimes\dk \equiv
|0\>\otimes\uk$. If $i$ is odd then in every amplitude the index $i$
needs to be replaced by $i-1$.

This generalization of the traditional coined quantum walks consists
in adding the possibility of self-transitions for each site. One
can, in principle, consider even more general transitions. As long
as the evolution is governed by a unitary operator with a cyclic
vector there is a CMV matrix lurking around. In our case the basis
is given directly in terms of the basic states and there is no need
to look for a new basis.

It is clear that the main results in \cite{CGMV} extend to this more
general case. If we have a way of computing the orthogonal Laurent
polynomials we get an integral expression for the transition
amplitude for going between any pair of basic states in any number
of steps.

\section{The Schur function for Riesz's measure}
From the expression for the Riesz measure given earlier we see that
the expansion
\[
d\mu(z)  = \sum_{j=-\infty}^{\infty} \overline\mu_j z^j
\frac{dz}{2\pi iz}
\]
leads to the moments $\mu_j$ of the measure $\mu$. Apart form the
first one, $\mu_0=1$, if $j\neq0$ can be written, in the necessarily
unique form, as
\[
j= \pm 4^{k_1} \pm 4^{k_2} \pm \dots \pm 4^{k_p}, \qquad k_1> k_2 >
\dots >k_p \ge 1,
\]
then
\[
\mu_j=1/2^p.
\]
For values of $j$ that cannot be written in the form above we have
$\mu_j=0$. In particular for $j=4^k$ we have $\mu_j=1/2$.

The moments of $\mu$ provide the Taylor expansion of the
Carath\'{e}odory function
\[
F(z)=1+ 2 \sum_{j=1}^{\infty} \overline\mu_j z^j,
\]
and from this it is not hard to compute the first few terms of the
Taylor expansion of the Schur function $f(z)$ around $z=0$. Indeed,
from
\[
F(z)= 1 + z^4 + \frac{z^{12}}{2} + z^{16} + \frac{z^{20}}{2} +
\frac{z^{44}}{4} + \frac{z^{48}}{2} + \frac{z^{52}}{4} +
\frac{z^{60}}{2} + z^{64} + \cdots
\]
we get that $f(z)$ has the expansion
\[
f(z) = \frac{z^3}{2} - \frac{z^7}{4} + \frac{3z^{11}}{8} +
\frac{3z^{15}}{16} - \frac{z^{19}}{32} - \frac{5z^{23}}{64} -
\frac{17z^{27}}{128} - \frac{29z^{31}}{256} + \cdots\,.
\]

Only powers differing in multiples of 4 appear in the Taylor
expansion of both functions, $F$ and $f$. This follows from the fact
that $d\mu(z)=d\nu(z^4)$ with $\nu$ given by the same infinite
product (\ref{RIESZ}) as $\mu$ but starting at $k=0$, i.e.
$d\nu(z)=(1+ (z+z^{-1})/2)\;d\mu(z)$. From this we find that
$F(z)=G(z^4)$ and $f(z)=z^3g(z^4)$ where $G$ and $g$ are the
Carath\'{e}odory and Schur functions of $\nu$ respectively.

It is now possible, in principle, to compute as many Verblunsky
parameters for the function $g$ as one wishes; they are given by the
continued fraction algorithm given at the end of section \ref{CMV}.
The first few ones are given below, arranged for convenience in
groups of eight. We list separately the first four parameters.
\[
\begin{array}{rrrrrrrr}
& & & & 1/2 &-1/3& 5/8& -1/13\\
1/14& -1/15& -1/4 &-1/9 &1/10 &-1/11 &21/32& -1/53\\
1/54& -1/55& -3/52 &-1/49& 1/50 &-1/51&5/56 &-1/61\\
1/62& -1/63 &-1/20 &-1/57& 1/58 &-1/59 &-11/48& -1/37\\
1/38& -1/39 &-1/12 &-1/33 & 1/34 &-1/35 &1/8 &-1/45\\
\dots
\end{array}
\]
It is clear that we get the Verblunsky parameters of $f$ by
introducing three zeros in between any two values above (a
consequence of the argument $z^4$ in $g$ above) and then shifting
the resulting sequence by adding three extra zeros at the very
beginning (a consequence of the factor $z^3$ in front of $g$ above),
yielding finally the following sequence of Verblunsky parameters for
$f$, where each row contains eight coefficients starting with
$\alpha_0$ in the first row, $\alpha_8$ in the second one, etc.
\[
\begin{array}{rrrrrrrr}
0&0&0&1/2&0&0&0&-1/3\\
0&0&0&5/8&0&0&0&-1/13\\
0&0&0&1/14&0&0&0&-1/15\\
0&0&0&-1/4&0&0&0&-1/9\\
0&0&0&1/10&0&0&0&-1/11\\
0&0&0&21/32&0&0&0&-1/53\\
0&0&0&1/54&0&0&0&-1/55\\
0&0&0&-3/52&0&0&0&-1/49\\
0&0&0&1/50&0&0&0&-1/51\\
0&0&0&5/56&0&0&0&-1/61\\
\dots
\end{array}
\]

The non-zero Verblunsky parameters of $f$ are given by
\[
\xi_m \equiv \alpha_{4m-1}, \qquad m=1,2,3,\dots
\]
where the sequence $\{\xi_m\}$ will be determined below. In fact it
will be enough to determine the subsequence $\{\xi_{4+8n}\}$ since
all the other values of $\xi_m$ can be given by simple expressions
in terms of these.

The expression for these $\xi_{4+8n}$ $\equiv$ $\alpha_{15+32n}$ is
given by
\[
\alpha_{15+32n}=-1/A_{n+1}, \qquad n=0,1,2,\dots
\]
for a sequence of integer values $\{A_n\}$ to be described below. We
will later give a different description of the complete sequence
$\{\xi_m\}$ which makes clear what its limit points are and obviates
the need to consider the subsequence $\{\xi_{4+8n}\}$.

We will first describe a procedure that allows us to generate the
infinite sequence of integers $\{A_n\}_{n=1}^\infty$ of which the
first ones are
\[
\begin{matrix}
13,53,61,37,45,213,221,197,205,245,253,229,237,149,157,133,141,\dots
\end{matrix}
\]
Once this sequence is accounted for, i.e. if the Verblunsky
parameters of the form $\alpha_{15+32 n}$ are known, we will see
that all the remaining ones are determined by simple explicit
formulas in terms of $\{A_n\}$. For this reason we will refer to the
sequence $\{A_n\}$ to be constructed in the next section as the
backbone of the sequence $\{\alpha_n\}$ we are interested in.

As we noted at the end of introduction, F.~Riesz included the factor
corresponding to $k=0$ in the infinite product (\ref{RIESZ}). That
is, the measure considered by Riesz is the measure $\nu$ giving our
measure $\mu$ (starting the infinite product with $k=1$) by
replacing $z$ by $z^4$, and the corresponding Schur function is $g$.
Therefore, the Verblunsky parameters $\{\alpha_n^R\}$ that F.~Riesz
would have are those obtained deleting in the sequence
$\{\alpha_n\}$ the groups of three consecutive zeros, so
$\alpha_n^R=\alpha_{4n+3}=\xi_{n+1}$, and all of them should be
computed from $\alpha^R_{8n-5}=-1/A_n$.

The main difference between including the factor $k=0$ in
(\ref{RIESZ}) or leaving it out is the inclusion of many zeros in
the list of Verblunsky parameters in the second case, which is the
one we choose. This makes for a much sparser CMV matrix which is
easier to analyze than it would be in the original case of F.~Riesz.
On the other hand his choice is better for computational purposes
when, of necessity, one has to deal with truncated matrices. In
Riesz's case there is more information packed in the same size
finite matrix. This point is exploited in some of the graphs
displayed at the end of the paper.

\section{Building the backbone}
Consider the sets $v_j$, $j \ge 0$, defined as the ordered set of
integers of the form
\[
((-2)^j-1)/3+k 2^{j+1}
\]
where $k$ runs over the integers. As an illustration we give a few
elements of the sets $v_0,v_1,v_2,v_3,\dots,v_{10}$, namely
\[
\begin{array}{l}
v_0=\dots,-6,-4,-2,0,2,4,6,\dots;\\ v_1=\dots,-13,-9,-5,-1,3,7,11,15,\dots;\\
v_2=\dots,-23,-15,-7,1,9,17,25,\dots; \\
v_3=\dots,-35,-19,-3,13,29,45,\dots;\\ v_4=\dots,-91,-59,-27,5,37,69,101,\dots;\\
v_5=\dots,-267,-203,-139,-75,-11,53,117,181,\dots; \\
v_6=\dots,-491,-363,-235,-107,21,149,277,405,\dots;\\ v_7=\dots,-1067,-811,-555,-299,-43,213,469,725,\dots;\\
v_8=\dots,-1963,-1451,-939,-427,85,597,1109,\dots; \\
v_9=\dots,-2219,-1195,-171,853,1877,2901,\dots;\\
v_{10}=\dots,-5803,-3755,-1707,341,2389,4437,\dots .
\end{array}
\]

These sets $v_j$ , $j \ge 0$, are disjoint and their union gives all
integers. A simple argument to prove this was kindly supplied by
B.~Poonen.

We observe that $d_j$, defined as the first positive element in the
infinite set $v_j$ (corresponding either to the choice $k=0$ or
$k=1$) is given as follows: if $j=0$ then $d_0=2$ otherwise, for $n
\ge 1$ we have
\[
d_j= ((-2)^j - 1)/3 + (1-(-1)^j) 2^j.
\]
Define now, for $n \ge 4$,
\[
c_n= 8+((-2)^{n-4}-1) 2^{5}/3
\]
so that the values of $c_4,c_5,c_6,c_7,\dots$ are given by
$$8, -24, 40, -88, 168, -344, 680, -1368, 2728,\dots$$
a sequence whose first differences are given by
\[
(-2)^{n+1},\quad n=4,5,6,\dots\,.
\]
For each pair $j,n$, $j \ge 0$, $n \ge 0$, define $w_{j,n}$ as the
number of positive elements in the sequence $\{v_j\}$ that are not
larger than $n$. Notice that $\sum_{j=0}^{\infty} w_{j,n}=n$ and
that $w_{j,0}$ is zero for all $j.$

To be very explicit, we have for instance $w_{0,40}=20$,
$w_{1,40}=10$, $w_{2,40}=5$, $w_{3,40}=2$, $w_{4,40}=2$,
$w_{5,40}=0$, $w_{6,40}=1$, and all values of $w_{j,40}$ after these
ones vanish.

It is possible to give an expression for $w_{j,n}$ in terms of the
sequence $\{d_j\}$ defined above. In fact one has
\[
  w_{j,n} =  \left\lfloor  \frac{n+2^{j+1}-d_{j}}{2^{j+1}} \right\rfloor
\]
where we use the notation $\lfloor x \rfloor$  to indicate the
integer part of the quantity $x.$

We are finally ready to put all the pieces together and define, for
$n \ge 0$,
\begin{equation}
u_n=\sum_{j=0}^{\infty} w_{j,n} c_{4+j}.
\end{equation}
Notice that by definition this is a finite sum since, for a given
$n$ the expression $w_{j,n}$ vanishes if $j$ is large enough.

The reader will have no difficulty verifying that we get
\begin{equation}
A_i=13 + u_{i-1}, \qquad i \ge 1.
\end{equation}
Recall that we have, for $n \ge 0$,
\begin{equation}
\alpha_{15+32 n}= -\frac{1}{A_{n+1}}
\end{equation}
and, as mentioned earlier, we will see how all other values of
$\alpha_j$ can be determined from these ones.

\section{Building the sequence from its backbone}
We have observed already that the first non-zero Verblunsky
parameters of our measure are given by
\[
\alpha_3=1/2,\quad \alpha_7=-1/3, \quad \alpha_{11}=5/8, \quad
\alpha_{15}=-1/13, \; \dots\,.
\]
We will see now that, for $i \ge 15$ and using the sequence
$\{A_i\}$ built above, we have a way of computing the non-zero
values of $\alpha_j$. Start by observing that, after
$\alpha_{15}=-1/A_1$ we get for values of $j$ between $j=16$ and
$j=47$ the following non-zero Verblunsky parameters:
\[
\alpha_{19}=\frac{1}{1+A_1}, \qquad \alpha_{23}=-\frac{1}{2+A_1},\
\]
followed by
\[
\alpha_{31}=-\frac{1}{A_1-4},\qquad \alpha_{35}=\frac{1}{A_1-3},
\qquad \alpha_{39}=-\frac{1}{A_1-2},\
\]
and finally
\[
\alpha_{47}=-\frac{1}{A_2}\,.
\]

The reader will have noticed that we did not give a prescription for
$\alpha_{27}$ or for $\alpha_{43}$.  This is done now:
\[
\alpha_{27}=-\frac{3}{A_1-1}, \qquad
\alpha_{43}=\frac{A_2-A_1+2}{A_2+A_1-2}\,.
\]
We have seen that the non-zero values of $\alpha_j$ for $j$ in
between $j=15$ and $j=47$ are all obtained from the values of $A_1$
and $A_2$. We claim that exactly the same recipe apply for values of
$j$ in the range from $16(2 p-1) -1$ and $16 (2 p+1)-1$, with $p \ge
1$, namely we set
\[
\alpha_{16 (2p-1)-1}=-\frac{1}{A_{p}},\qquad
\alpha_{16(2p+1)-1}=-\frac{1}{A_{p+1}}
\]
and fill in the {\bf SEVEN} non-zero values of $\alpha_j$ in between
these two by using expressions that are extensions of the ones
above, namely,
\[
\begin{aligned}
& \alpha_{16(2p-1)+3}=\frac{1}{1+A_p}, \qquad & &
\alpha_{16(2p-1)+7}=-\frac{1}{2+A_p},
\\
& \alpha_{16(2p-1)+15}=-\frac{1}{A_p-4}, \qquad & &
\alpha_{16(2p-1)+19}=\frac{1}{A_p-3},
\\
& \alpha_{16(2p-1)+23}=-\frac{1}{A_p-2}, \qquad & &
\alpha_{16(2p-1)+31}=-\frac{1}{A_{p+1}},
\end{aligned}
\]
and, just as above, the missing Verblunsky parameters are given by
\[
\alpha_{16 (2p-1)+ 11}=-\frac{3}{A_p-1},\qquad \alpha_{16
(2p-1)+27}=\frac{A_{p+1}-A_p+2}{A_{p+1}+A_p-2}\,.
\]

\section{A different expression for the Verblunsky parameters}
The construction above gives as many non-zero Verblunsky parameters
for our measure as one wants, starting with
$\alpha_3,\alpha_7,\alpha_{11},\alpha_{15},\dots$ for which we get
the values $1/2,1/3,5/8,-1/13,\dots$\,.

In this section we give an explict formula for these Verblunsky
parameters in terms of a sequence of constants $\{K_i\}$,
$i=0,1,2,\dots$, which are closely related to the sequence $\{A_i\}$
introduced above. One of the advantages of this new expression is
that the set of limit values of the sequence $\{\alpha_j\}$ becomes
obvious and is given by the union of three infinite sets, namely
\[
\begin{aligned}
-\frac{2}{K_i}, & & \qquad & i=1,2,3,\dots \, ,
\\
\frac{4}{K_{i}+3}, & & \qquad & i=0,1,2,3,\dots \, ,
\\
-\frac{2}{K_{i}+6}, & & \qquad & i=0,1,2,3,\dots \, .
\end{aligned}
\]
where the constants $K_i$ are given by
\[
\begin{aligned}
& K_0=3, \\ & K_{2i-1}=3 A_i, \qquad K_{2i}= 3(A_i-4), \qquad
i=1,2,3,\dots\,.
\end{aligned}
\]
One needs to add the limit points of the three infinite sets given
above to get all limit points of the sequence $\{\alpha_j\}$.

We are now ready to give the alternative expressions for the
non-zero Verblunsky parameters alluded to above, i.e. $\xi_n$ so
that
\[
\xi_1=1/2, \quad  \xi_2=-1/3, \quad \xi_3=5/8, \; \dots \, ,
\]
and in general $\alpha_{4m-1}=\xi_m.$ For this purpose we consider a
disjoint union of the set of all non-negative integers into sets
$B_n$, where the index $n$ runs over the set
${1,2,4,5,6,8,\dots}$\,, i.e. all positive $n\ne 3\pmod 4.$

For each such $n$, define $B_n$ as the set of integers of the form
\[
\frac{1}{3} + 4^p \frac{3n-1}{3}, \qquad p=0,1,2,3,\dots\,.
\]
Once again, a simple proof of these properties of the sets $B_n$ was
supplied by B.~Poonen.

The sets $B_n$ break naturally into three classes, with $n\equiv
0\pmod 4$, $n\equiv 1\pmod 4$ and $n\equiv 2\pmod 4$. We claim that
\[
\begin{aligned}
& \xi_{\frac{1}{3}+4^p\frac{3n-1}{3}} =
-\frac{1}{K_s}\left(2+\frac{1}{4^p}\right), \qquad n=4s, \qquad
s=1,2,3,\dots \, ,
\\
& \xi_{\frac{1}{3}+4^p\frac{3n-1}{3}} =
\frac{1}{K_s+3}\left(4-\frac{1}{4^p}\right), \qquad n=4s+1, \qquad
s=0,1,2,\dots\, ,
\\
& \xi_{\frac{1}{3}+4^p\frac{3n-1}{3}} =
-\frac{1}{K_s+6}\left(2+\frac{1}{4^p}\right), \qquad n=4s+2, \qquad
s=0,1,2,\dots\,.
\end{aligned}
\]

From the expressions above it follows that we have identified the
limit values of the sequence $\{\alpha_{4m-1}\}$. The largest one is
$2/3$ and the lowest one $-2/9$.

\section{Some properties of the Riesz quantum walk}
Once we get our hands on the Verblunsky parameters corresponding to
the Riesz measure we construct the corresponding CMV matrix and we
can compute different quantities pertaining to the associated
quantum walk. In the rest of the paper we choose to illustrate some
of these results with a few plots.

\begin{figure}
  \vspace*{-12cm}
  \includegraphics[height=30.5cm,width=12cm,angle=0]{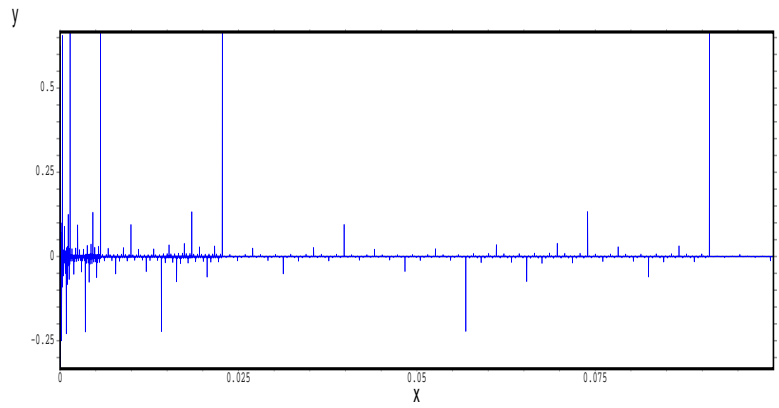}
  \caption{Riesz's measure: The first 30000 non-zero Verblunsky parameters.}\label{Fi:nice}
\end{figure}

\begin{figure}
     \vspace*{-12cm}
     \includegraphics[height=30.5cm,width=12cm,angle=0]{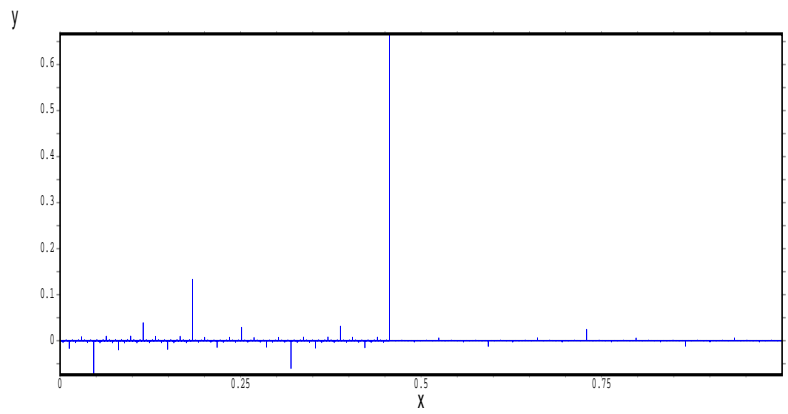}
     \caption{Riesz's measure: The non-zero Verblunsky parameters for indices between
     30000 and 60000.}\label{Fi:very nice}
\end{figure}

Figures~\ref{Fi:nice} and \ref{Fi:very nice} display the Verblunsky
parameters themselves, key ingredients in the one-step transition
amplitudes of the Riesz quantum walk. They show an apparent chaotic
behaviour which is actually driven by the rules previously described
which generate the full sequence of Verblunsky parameters. This is
in great contrast to the translation invariant case of a constant
coin.

\begin{figure}
   \vspace*{-12cm}
   \includegraphics[height=30.5cm,width=12cm,angle=0]{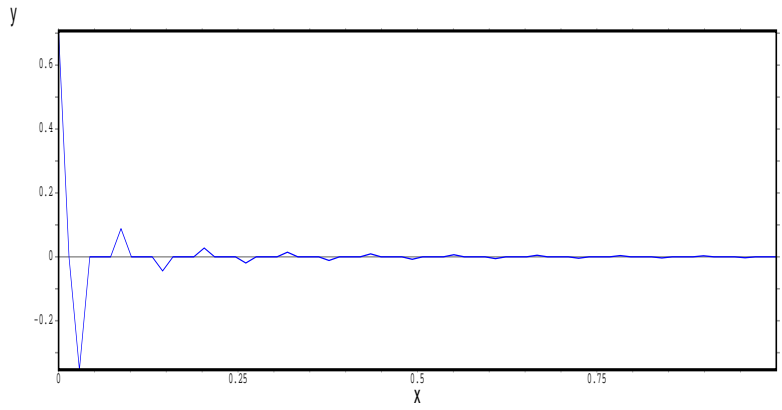}
   \caption{Hadmard's walk: The first 70 non-zero Taylor coefficients of
   the Schur function.}\label{Fi:uglyhad}
\end{figure}

\begin{figure}
   \vspace*{-12cm}
   \includegraphics[height=30.5cm,width=12cm,angle=0]{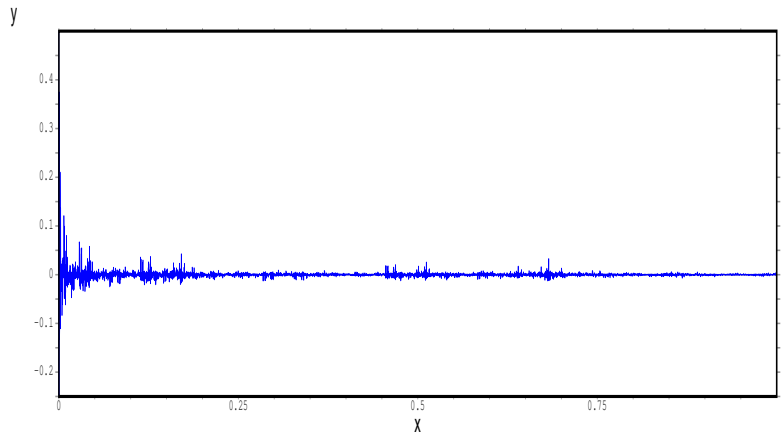}
   \caption{Riesz's measure: The first 7000 non-zero Taylor coefficients of
   the Schur function.}\label{Fi:ugly}
\end{figure}

Figures~\ref{Fi:uglyhad} and \ref{Fi:ugly} display the Taylor
coefficients of the Schur function for Riesz measure and the
Hadamard quantum walk with constant coin
\[
C = \frac{1}{\sqrt{2}} \begin{pmatrix} 1 & 1 \\ 1 & -1 \end{pmatrix}
\]
on the non-negative integers. These coefficients have an important
probabilistic meaning discussed in great detail in the forthcoming
paper \cite{GVWW}: the $n$-th Taylor coefficient is the first time
return amplitude in $n$ steps to the state with spin up at site 0.

\begin{figure}
   \vspace*{-12cm}
   \includegraphics[height=30.5cm,width=12cm,angle=0]{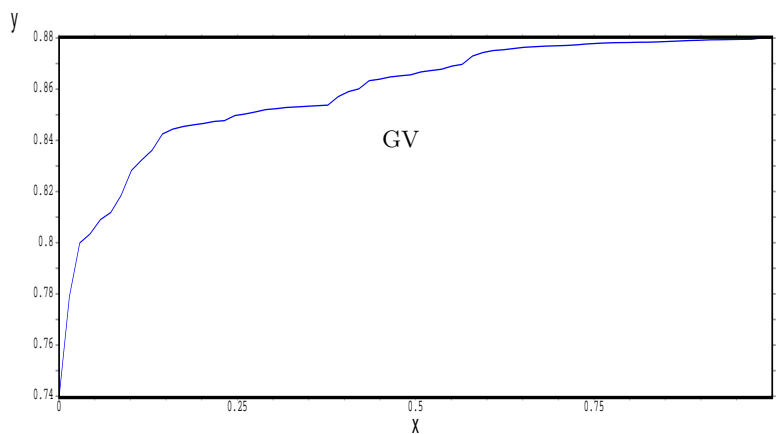}
   \caption{Riesz's measure: The cumulative sums of the squares of the first 7000
   non-zero Taylor coefficients of the Schur function in steps of 100.}\label{Fi:soso}
\end{figure}

The first time return amplitudes for the Riesz walk seem to
fluctuate in an apparent random way around a mean value which must
decrease strongly enough to ensure that the sequence is
square-summable, as Figure~\ref{Fi:soso} makes evident. This is
because the sum of the first return probabilities is the total
return probability, which cannot be greater than one. Equivalently,
any Schur function is Lebesgue integrable on the unit circle with
norm bounded by one, and its norm is the sum of the squared moduli
of the Taylor coefficients.

\begin{figure}
   \vspace*{-12cm}
   \includegraphics[height=30.5cm,width=12cm,angle=0]{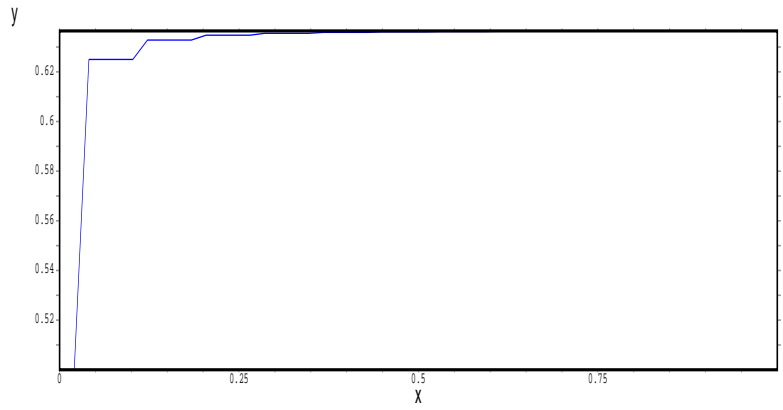}
   \caption{Hadamard's walk: The cumulative sums of the squares of the first 70
   non-zero Taylor coefficients of the Schur function.}\label{Fi:sosohad}
\end{figure}

In the Hadamard case the behaviour of the first time return
amplitudes is much more regular and the convergence to the total
return probability, depicted in Figure~\ref{Fi:sosohad}, holds with
a much higher speed. We should remark that the plot for the Hadamard
walk picks up only the first 70 non-null coefficients, while the
Riesz picture represents the first non-null 7000 coefficients.
Hence, the differences between these two examples are not only in
the more regular pattern that the Hadamard return probabilities
exhibit, but also in the much higher tail for the Riesz return
probabilities.

\begin{figure}
   \vspace*{-12cm}
   \includegraphics[height=30.5cm,width=12cm,angle=0]{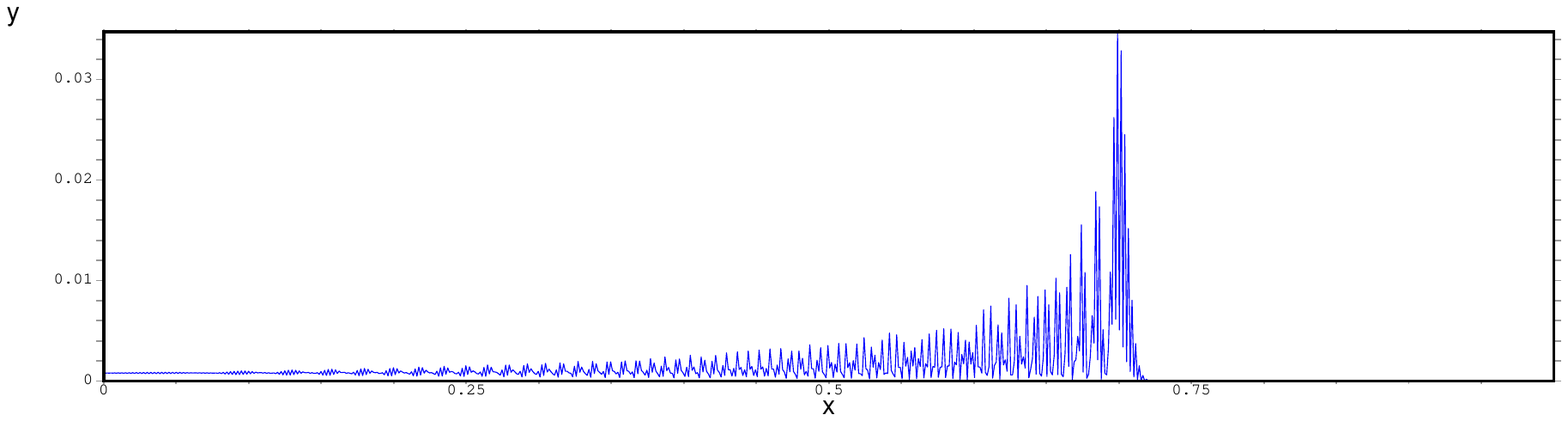}
   \caption{Hadamard's walk on the non-negative integers: The probability of
   $X_n/n$ for 800 iterations starting at $|0\>\otimes\uk$.}\label{Fi:ops}
\end{figure}

\begin{figure}
   \vspace*{-12cm}
   \includegraphics[height=30.5cm,width=12cm,angle=0]{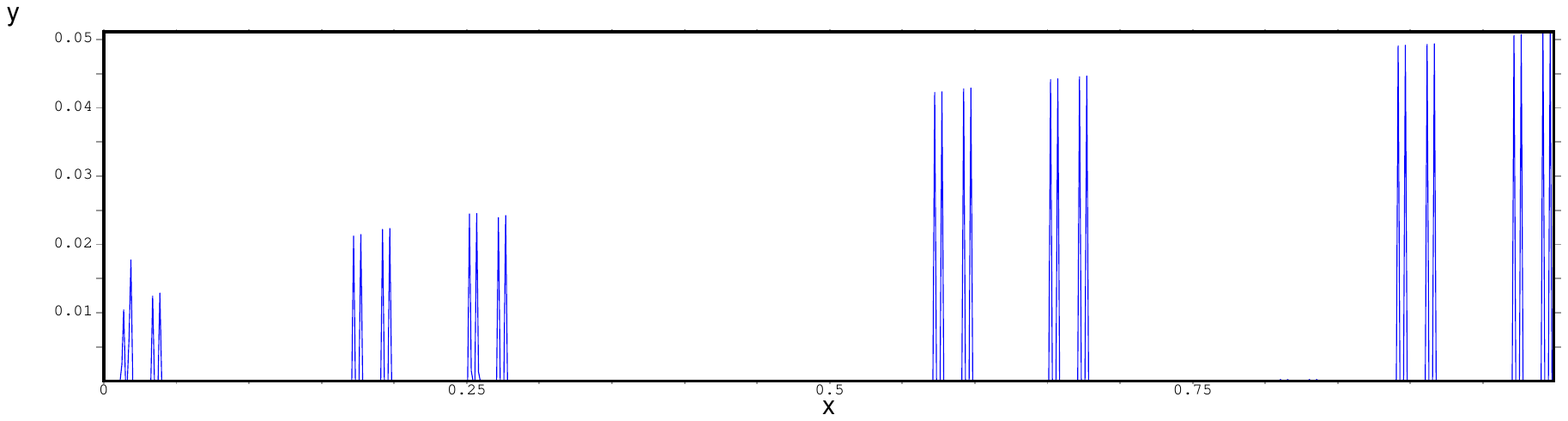}
   \caption{Riesz's walk on the non-negative integers: The probability of
   $X_n/n$ for 800 iterations starting at $|0\>\otimes\uk$.}\label{Fi:oops}
\end{figure}

Figures~\ref{Fi:ops} and \ref{Fi:oops} give the probability
distribution of the random variable $X_n/n$, where $X_n$ stands for
the position (regardless of spin orientation) after $n$ steps of the
quantum walk started at position $0$ with a spin pointing up. The
plots given here correspond to the value $n=800$ for both, the Riesz
quantum walk and the Hadamard constant coin on the non-negative
integers.

The figures show that, in contrast to classical random walks for
which $X_n$ behaves tipically as $\sqrt{n}$, the position in a
quantum walk can grow linearly with $n$. Nevertheless,
Figure~\ref{Fi:oops} shows a striking behaviour of the Riesz walk
compared to the more regular asymptotics of the Hadamard walk
reflected in Figure~\ref{Fi:ops}. This should be viewed as a clear
indication of the anomalous behaviour that can appear under the
presence of a singular continuous spectrum. In particular, these
results make evident that quantum walks with a singular continuous
measure can not exhibit nice limit laws as other toy models do. For
the case of translation invariant ones it is known that obey
inverted bell asymptotic distributions (see for instance \cite{Ko}).

These results should motivate a more detailed analysis of quantum
walks associated with singular continuous measures. This could lead
to the discovery of new interesting quantum phenomena.


\end{document}